\RequirePackage{amsmath}
\documentclass[runningheads, envcountsame, a4paper]{llncs}

   \usepackage[T1]{fontenc}
   \usepackage{graphicx}
   \usepackage{tikz}
   \usepackage{environ}
   \usepackage{pgfplots}
   \usepackage{multirow}
   \usepackage{booktabs}
   \usepackage{float}
   \usepackage{bm}
   \usepackage{microtype}
   \usepackage{filecontents}
   \usepackage[misc]{ifsym}
   \usepackage[hyphens]{url}
   \usepackage[linesnumbered,ruled]{algorithm2e}
   \usepackage[group-separator={,}]{siunitx}

   \definecolor{myblue}{RGB}{81,115,213}

\begin{document}
   \pagestyle{plain}

   \title{
      Marine Mammal Species Classification using Convolutional Neural Networks and a Novel Acoustic Representation
   }
   \toctitle{
      Marine Mammal Species Classification using Convolutional Neural Networks and a Novel Acoustic Representation
   }
   \author{
      {Mark Thomas \inst{1}\Letter} \and
      Bruce Martin \inst{2} \and
      Katie Kowarski \inst{2} \and
      Briand Gaudet \inst{2} \and
      {Stan Matwin \inst{1,3}}\thanks{Stan Matwin's research is supported by the Natural Sciences and Engineering Research Council and by the Canada Research Chairs program.}
   }
   \tocauthor{
      Mark~Thomas, Bruce~Martin, Katie~Kowarski, Briand~Gaudet, and Stan~Matwin
   }
   \institute{
      Dalhousie University Faculty of Computer Science, Halifax, Canada \\ \email{mark.thomas@dal.ca,stan@cs.dal.ca} \and
      JASCO Applied Sciences, Dartmouth, Canada \\ \email{$\{$bruce.martin,katie.kowarski,briand.gaudet$\}$@jasco.com} \and
      Institute of Computer Science Polish Academy of Sciences, Warsaw, Poland
   }
   \maketitle
   \setcounter{footnote}{0}
   \begin{abstract}
      Research into automated systems for detecting and classifying marine mammals in acoustic recordings is expanding internationally due to the necessity to analyze large collections of data for conservation purposes. In this work, we present a Convolutional Neural Network that is capable of classifying the vocalizations of three species of whales, non-biological sources of noise, and a fifth class pertaining to ambient noise. In this way, the classifier is capable of detecting the presence and absence of whale vocalizations in an acoustic recording. Through transfer learning, we show that the classifier is capable of learning high-level representations and can generalize to additional species. We also propose a novel representation of acoustic signals that builds upon the commonly used spectrogram representation by way of interpolating and stacking multiple spectrograms produced using different Short-time Fourier Transform (STFT) parameters. The proposed representation is particularly effective for the task of marine mammal species classification where the acoustic events we are attempting to classify are sensitive to the parameters of the STFT.
   \end{abstract}

%% Introduction -------------------------
\section{Introduction}\label{sec:introduction}
   Since their introduction to the area of computer vision, Convolutional Neural Networks (CNNs) have continued to improve upon the state-of-the-art. Recently, a growing collection of research has been brought forward applying CNNs to tasks which are auditory in nature, including: speech recognition \cite{abdel2014,deng2013}, musical information retrieval \cite{choi2017,humphrey2012}, and acoustic scene classification \cite{piczak2015,salamon2016}.

   Inspired by the compelling results obtained in the previously mentioned domains, researchers in oceanography and marine biology have started to investigate similar solutions to problems in their field. One such problem is the analysis of underwater acoustic data, which is one of the primary methods used to measure the  presence, abundance, and migratory patterns of marine mammals \cite{zimmer2011}. The necessary acoustic data for modelling marine mammal life is often collected using Passive Acoustic Monitoring (PAM) techniques. PAM is non-invasive and reduces the risk of altering the behaviour of a species of interest, unlike GPS tagging. PAM is also less susceptible to harsh weather conditions compared to visual surveys. Acoustic data collected for PAM is often carried out using moored recorders equipped with hydrophones. Stakeholders make use of PAM to adjudicate environmental and governmental policy decisions, for example implementing reduced speed limits on vessels travelling through shipping channels in order to reduce their risk of collision with endangered species of whales \cite{shipping2019}.

   Due to their high cost of deployment, PAM recording devices may be left unattended for months or years at a time before resurfacing, producing very large amounts of data; typically several terabytes per deployment. It is becoming increasingly common for collections of acoustic data to be described at the petabyte scale, making complete human analysis infeasible. As a result, research into automated Detection and Classification Systems (DCS) is widespread and continuing to grow. From a machine learning perspective, a DCS can be interpreted as a hierarchical model containing a binary classifier recognizing whether a signal of interest is present within an acoustic recording, combined with a multi-class classifier for determining the source of the signal. Importantly, marine biologists and oceanographers are typically concerned with the presence or absence of specific species in an acoustic recording. While there have been great advances in the research and development of these systems, many DCS are based on the acoustic properties of a signal of interest and may be specific on a per-dataset basis depending on the equipment that was used or the geographic location of the recordings. Therefore, such systems are often not generalizable and may require being formulated from scratch for a new data set. Moreover, attempts at producing generalizable systems yield high rates of false detections \cite{baumgartner2011}.

   In this work, we present a deep learning implementation of a DCS composed of a CNN trained on spectrogram representations of acoustic recordings. The main contributions of this work are:

   \begin{itemize}
      \item A CNN capable of classifying three species of marine mammals as well as non-biological sources and ambient noise.
      \item The classifier makes up an automated DCS that is generalizable and can be adapted to include additional species that produce vocalizations below 1000Hz.
      \item A novel visual representation of acoustic data based on interpolating and stacking multiple spectrograms produced using distinct Short-time Fourier Transform parameters.
   \end{itemize}

   This work describes a complete application using original data collected for scientific research that could have substantial implications towards environmental policy and conservation efforts. The data was manually selected based on the target species of interest, however, it has not been cleaned and manipulated unlike many research projects in machine learning that use common sets of image data or preprocessed acoustic recordings. Additionally, while the results focused on in this paper are centred on detection and classification of marine mammals, the framework outlined in this paper can be adapted to other tasks such as acoustic scene classification.

   The remainder of this paper is organized as follows. In Section \ref{sec:background} we review related work on the topic of marine mammal species classification and provide further details on the complexities of the problem. An overview of common representations of acoustic data as well as a novel representation formulated especially for the task of marine mammal species classification is provided in Section \ref{sec:representations}. The data set used in training the CNN and additional information regarding the experimental setup is provided in Section \ref{sec:dataset}. The corresponding experimental results are analyzed in Section \ref{sec:results}. Finally, concluding remarks and future work are presented in Section \ref{sec:conclusion}.

%% Related Work -------------------------
\section{Background and Related Work}\label{sec:background}
   CNNs have traditionally been applied to visual recognition tasks on large collections of labelled images. Most notably, CNNs have lead to state-of-the-art performance for classifying commonly used benchmark image data sets and have surpassed human levels of performance \cite{he2016}. Beyond image classification, CNNs have also been used for object detection \cite{girshick2015,he2017} and in conjunction with Recurrent Neural Networks for natural language processing \cite{karpathy2015}.

   Recently, several factors have led researchers to apply CNNs outside of the visual paradigm such as classifying events or patterns found in acoustic recordings. An obvious reason for adapting CNNs to acoustic tasks is the performance levels of the classifiers cited above. A less obvious reason to those not working in the field of acoustics or digital signal processing, is that human analysis of acoustic data is often carried out visually using spectrograms as it is faster to visually identify signals of interest without having to listen to the entire recording. Another reason for using visual representations of acoustic data is that they allow for the analysis and interpretation of sounds outside of the human hearing range. One area, alluded to in Section \ref{sec:introduction}, that makes frequent use of visual representations of acoustic data is the detection and classification of marine mammal vocalizations within underwater acoustic recordings (i.e., DCS research).

   Research into automated DCS has been a growing topic of interest, in part, as a by-product of the reduced costs in recording equipment which has produced vast amounts of data. Another reason for the growth in DCS research is for conservation purposes, particularly as it relates to endangered species of whales. In developing an automated DCS for marine mammal vocalizations, one hopes to accurately detect and assign a label to an instance of an acoustic recording containing one or more vocalizations produced by a species of interest. However, developing a generalizable DCS presents several distinct challenges. For one, underwater recordings often have a low signal-to-noise ratio making feature extraction difficult. Another challenge is that ground truth labelled data is difficult to obtain due to the required expertise and training of the labeller. As a result, only a very small fraction of the large collections of acoustic data is suitable for supervised learning. Furthermore, the small numbers of some species coupled with the low rate of occurrence of their vocalizations make for highly unbalanced data.

   Traditionally, many of the algorithms used to detect and classify marine mammal vocalizations are derived from the properties of a signal of interest. In general, these approaches can be divided into two categories. The first category of algorithms involves comparing unlabelled data to templates of certain vocalizations. Examples of this approach include \textit{matched filtering}, where a template corresponding to the vocalization of interest is convolved with a signal to produce a detection function that is evaluated using a pre-determined threshold parameter \cite{clark1987}. Another example is \textit{spectrogram correlation}, which first computes a correlation kernel using segments of template spectrograms, following which, the correlation kernel is convolved over a spectrogram of the unlabelled data producing a vector representing the similarity between the spectrogram and the kernel over time. Large similarity values correspond to possible detections. The second category of algorithms involves detecting regions of interest in a spectrogram and extracting features (e.g.: the duration of the detection or the absolute change in frequency) to be used as input vectors for classification. Various detection algorithms are used in the first step of this approach including: neighbourhood search algorithms (e.g., pixel connectivity) in spectrograms that have been filtered, smoothed, and cast to binary representations \cite{baumgartner2011} and contour detectors that operate by continually searching for local maxima within pre-specified frequency bands of normalized spectra over time \cite{mellinger2011}. These detection algorithms are heavily dependent on the filtering, normalization, and smoothing operations that are performed on each spectrogram. Once the regions of interest are determined, feature vectors are then handed to commonly used classification algorithms such as: linear and quadratic discriminant analysis \cite{baumgartner2011,gillespie2013}, support vector machines \cite{dugan2010}, and artificial neural networks \cite{dugan2010}. Researchers have also likened the task to automatic speech recognition and used Gaussian mixture models and hidden Markov models for classification \cite{roch2011,skowronski2006}.

   The algorithms described above involve a large amount of human input--often from experts--which is a limitation to the development of future classifiers for several reasons. In the former category the templates used for detection and classification are largely specific to not only certain species, but also different types of vocalizations produced by the same species. Furthermore, the detection threshold may require fine-tuning depending on the noise characteristics of the data set. For the latter category of algorithms, many of the hyper-parameters provided to the smoothing and noise-removal routines are dependent on the data set. Subsequently, the hand-engineered features are contaminated by these specifications as well as human bias. These limitations yield systems which are not easily generalizable to a broad category of species using data collected at different sampling rates, geographic locations, or using different recording devices.

   More recently, researchers have attempted to use deep learning to learn generalizable representations of spectrograms for the purpose of DCS development. In one study, Halkias et al. \cite{halkias2013} contrast the performance of a restricted Boltzmann machine and a sparse auto-encoder for classifying five species of baleen whales (\textit{mysticetes}), however, the regions of interest containing the whale calls were assumed to be known. Wang et al. \cite{wang2018} use CNNs to classify spectrograms containing vocalizations of killer whales (\textit{Orcinus orca}) and pilot whales (\textit{Globicephala melas/macrorhynchus}) but similarly do not include non-biological or ambient noise sources. Liu et al. \cite{liu2018} also use CNNs but focus on the classification of call types as opposed to the species that produced them. Finally, Luo et al. \cite{luo2019} train a CNN to detect the  high-frequency echolocation clicks of toothed whales (\textit{odontocetes}) using a combination of real audio recordings and synthetic data, however, we are interested in classifying baleen whale vocalizations that occur at a much lower frequency and can be masked by low tonal sounds created by shipping activity.

%% Acoustic Representations -------------------------
\section{Visual Representations of Acoustic Data}\label{sec:representations}
   Human analysis of acoustic recordings is performed aurally by listening to an acoustic recording as well as visually using spectrograms. A popular approach for generating spectrograms is through a Short-time Fourier Transform (STFT). The STFT procedure calculates the sinusoidal frequency and phase content of an acoustic signal over time and is most commonly visualized in two dimensions with time on the $x$-axis, frequency on the $y$-axis, and intensity expressed by varying colour.

   The equation of the discrete-time STFT of a signal $x[n]$ can be expressed as:

      \begin{equation}\label{eqn:stft}
         X(n,\omega) = \sum_{m=-\infty}^\infty x[m]w[m-n]e^{-j\omega m}~,
      \end{equation}

   \noindent where $w$ is a windowing function with a pre-specified length centred at time $n$. In the equation expressed above, time is discrete and frequency ($\omega$) is continuous, however, in practice both units are discretized and each successive STFT is computed using an implementation of the Fast Fourier Transform (FFT) algorithm (e.g., the Cooley-Tukey algorithm \cite{cooley1965}). Equation \ref{eqn:stft} describes a complex function, therefore, we take the square of the absolute value of $X(n,\omega)$ yielding a spectrogram of the power spectral density. Finally, we convert the intensity from power to a logarithmic scale (i.e., decibels (dB)), as is commonly the case in underwater acoustics.

   \subsection{Mel-scaled Spectrograms}\label{subsec:mel-spectrograms}
      A spectrogram computed using the approach formulated above is linear in frequency. Unfortunately, because CNNs are spatially invariant, they are incapable of understanding human perceptions of pitch when frequency is expressed on a linear scale. For example, while the difference between two signals occurring at 1000Hz and 1500Hz and two other signals occurring at 10kHz and 10.5kHz are numerically equivalent (i.e., equal to 500Hz), the difference of the lower frequency signals is perceptually much larger to a human listener.

      The bandwidth of the data we are attempting to classify is relatively low (i.e., $\leq1000$Hz), therefore the CNNs imperception to pitch is not a major concern. However, in order to test this hypothesis, we additionally generate mel-scaled spectrograms whereby frequency is transformed from hertz to mels (from the word melody) using the formula outlined in Equation \ref{eqn:mels}.

         \begin{equation}\label{eqn:mels}
            \omega_{mel} = 2595\log_{10}\left(1 + \frac{\omega_{Hz}}{700}\right)~.
         \end{equation}

      Following this transformation, the resulting frequency scale more closely aligns with the $\log$-like human perception of pitch.

   \subsection{Novel Representation: Stacked \& Interpolated Spectrograms}\label{subsec:interpolated}
      The majority of the DCS detailed in Section \ref{sec:background} were trained using large collections of single channel inputs in the form of spectrograms. During the creation process of said data sets, a decision must be made on the appropriate combination of parameters to pass to the STFT. In practice, when marine biologists analyze acoustic recordings, they will often generate multiple spectrograms using different STFT parameters, for example: changing the length of the FFT window and/or the window overlap. By changing the parameters of the STFT, the time and frequency resolutions of the spectrogram are altered. Using multiple spectrograms with varying resolutions is particularly helpful when annotating underwater acoustic recordings containing marine mammal vocalizations because some species tend to make prolonged low-frequency vocalizations with a small bandwidth (e.g.: blue whale moans), while other species make shorter vocalizations with a larger bandwidth (e.g.: humpback songs). Depending on the set of parameters used to generate the spectrogram, one can easily misclassify a vocalization as a different species or miss the vocalization entirely.

         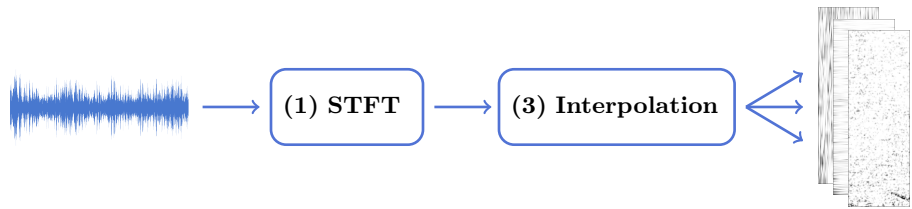
\begin{figure}[!htb]
            \centering
            \begin{tikzpicture}[
                  arrow/.style={->,myblue,line width=.2cm},
                  box/.style={myblue,rounded corners=.2cm,line width=.2cm}
               ]
               % Wav File
               \node[inner sep=0pt] at (-0.41,-0.05) {\includegraphics[width=2.5cm]{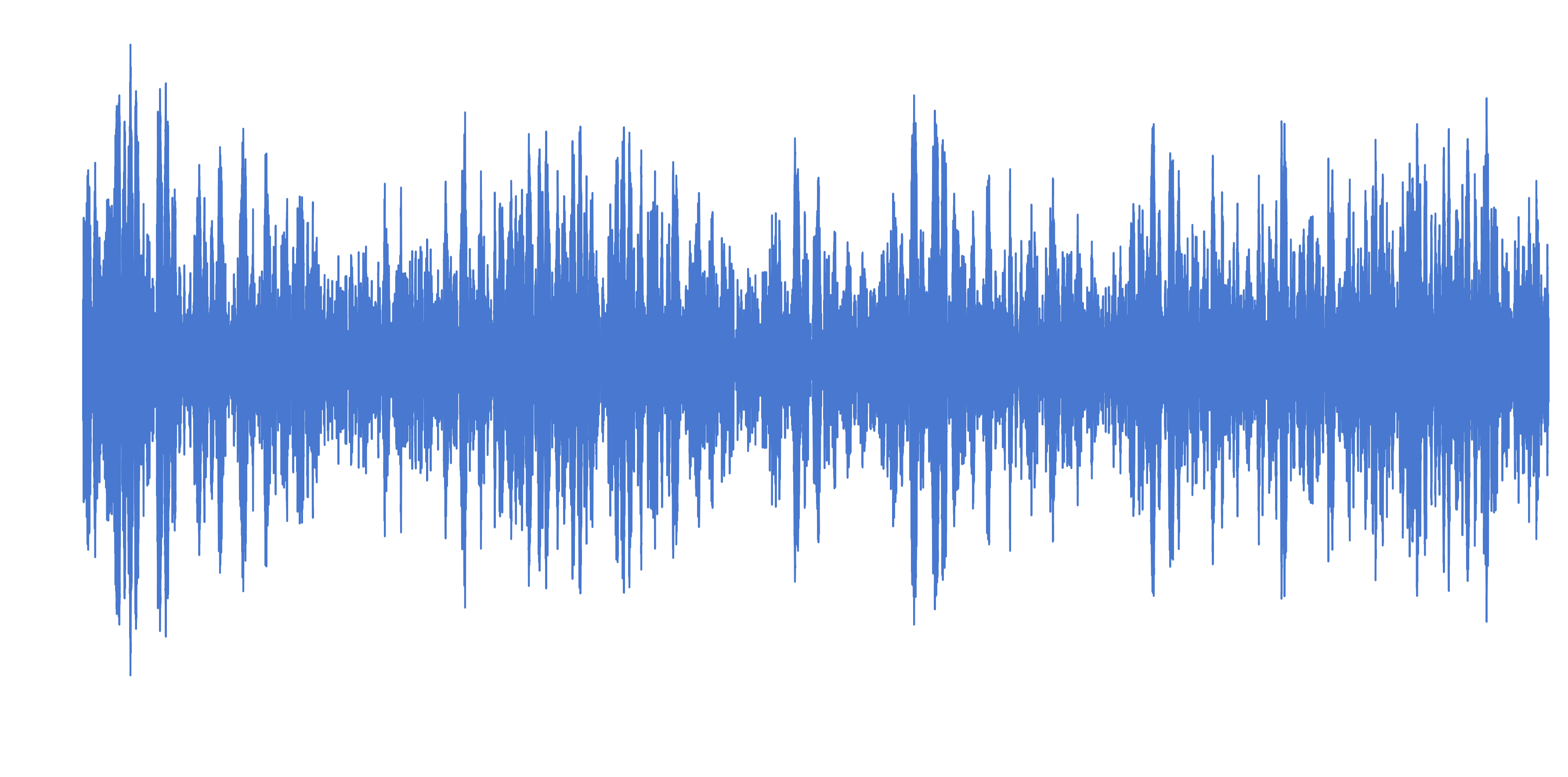}};
               \draw[arrow, line width=1pt] (1,0) -- (1.75,0) node{};

               % Box around STFT equation
               \pgfmathsetmacro{\cubex}{2};
               \pgfmathsetmacro{\cubey}{1};
               \draw[box, line width=1pt] (3.9,0.5,0) -- ++(-\cubex,0,0) -- ++(0,-\cubey,0) -- ++(\cubex,0,0) -- cycle;
               \node[text width=3cm] (fft) at (3.6,0) {\textbf{(1) STFT}};
               \draw[arrow, line width=1pt] (4.05,0) -- (4.8,0) node{};

               % Box around interpolation equation
               \pgfmathsetmacro{\cubex}{3.1};
               \pgfmathsetmacro{\cubey}{1};
               \draw[box, line width=1pt] (8,0.5,0) -- ++(-\cubex,0,0) -- ++(0,-\cubey,0) -- ++(\cubex,0,0) -- cycle;
               \node[text width=4cm] (fft) at (7.1,0) {\textbf{(3) Interpolation}};
               \draw[arrow, line width=1pt] (8.15,0) -- (8.9,0.45) node{};
               \draw[arrow, line width=1pt] (8.15,0) -- (8.9,0) node{};
               \draw[arrow, line width=1pt] (8.15,0) -- (8.9,-0.45) node{};

               % Original spectrograms
               \node[inner sep=0pt] at (9.5,0.15) {\includegraphics[width=.8cm]{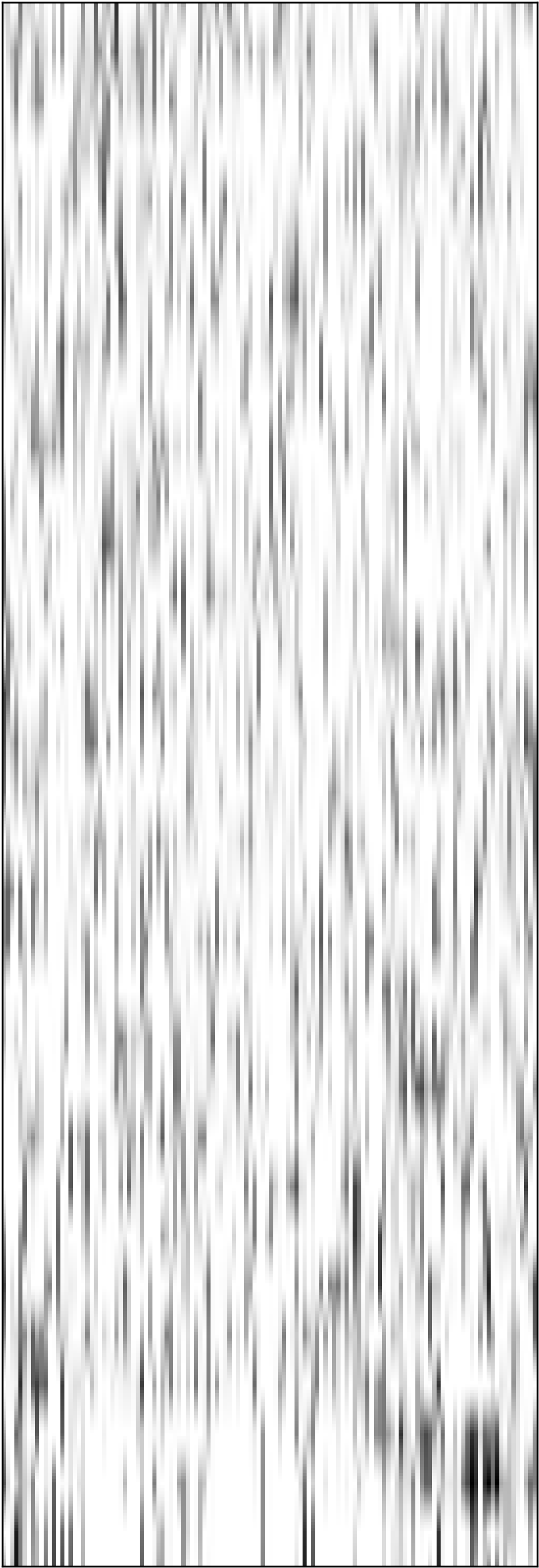}};
               \node[inner sep=0pt] at (9.7,0.00) {\includegraphics[width=.8cm]{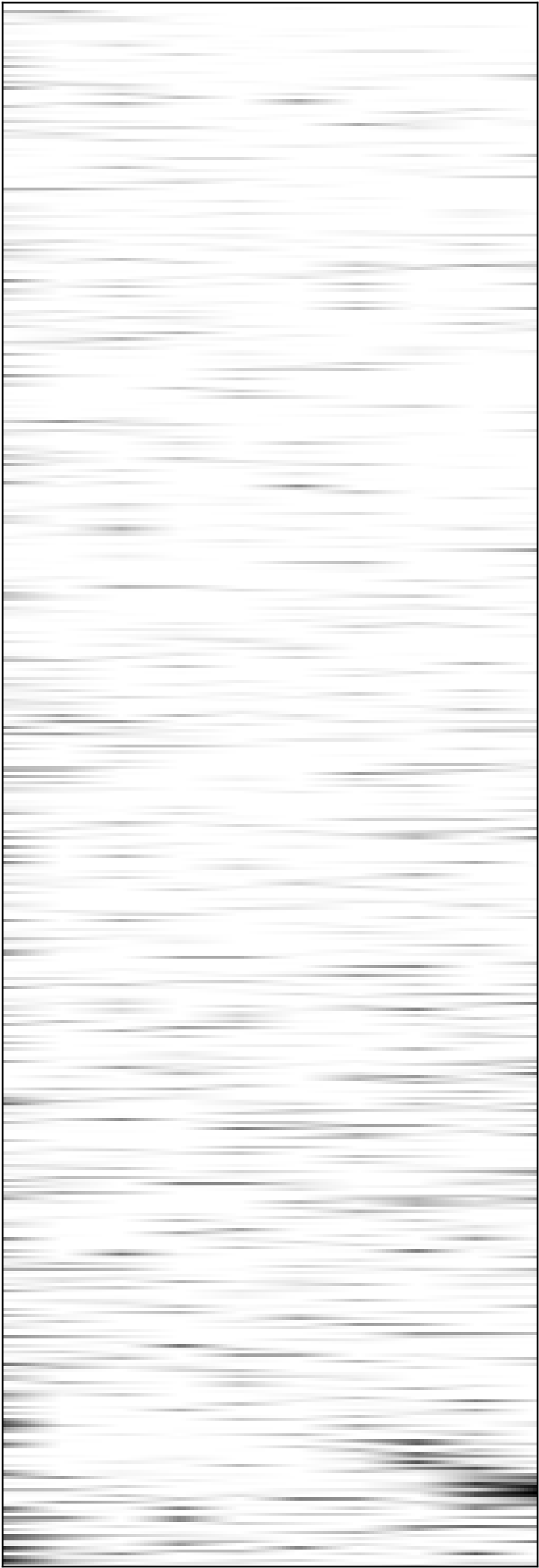}};
               \node[inner sep=0pt] at (9.9,-0.15) {\includegraphics[width=.8cm]{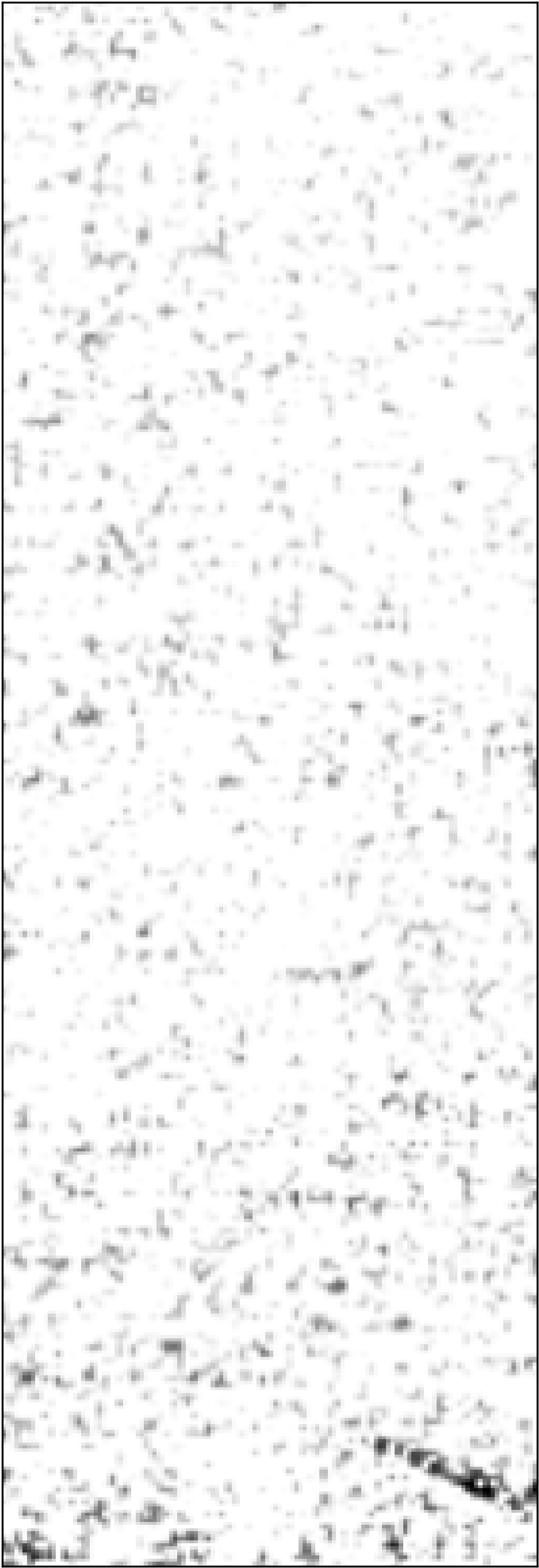}};
            \end{tikzpicture}
            \caption{Simple illustration demonstrating the process of transforming a waveform of an acoustic signal into a multi-channel input via interpolation and stacking.}
            \label{fig:data-pipeline}
         \end{figure}

      We propose a novel representation of an acoustic signal that attempts to exploit the strategy used by human experts during the annotation process. First, following Equation \ref{eqn:stft}, several spectrograms are generated using multiple sets of STFT parameters. Because each of the spectrograms vary in resolution across time and frequency, they are interpolated using a simple linear interpolation spline over a grid proportionate to the smallest time and frequency resolutions. The equation of a linear interpolation spline for some point $(n, \omega)$ between $(n_i, \omega_i)$ and $(n_{i+1}, \omega_{i+1})$, where $n$ is known, can be expressed as:

         \begin{equation}\label{eqn:interpolation}
            \omega = \omega_i + \frac{\omega_{i+1} - \omega_i}{n_{i+1} - n_i}(n - n_i)~.
         \end{equation}

      After interpolation, the dimensions of the matrices corresponding to each spectrogram are the same. The interpolated spectrograms are then stacked to form a multi-channel tensor; imitating the concept of RGB channels in a digital colour image, as depicted in Figure \ref{fig:data-pipeline}. The details of the algorithm used to produce a single instance of the novel representation described above are outlined in Algorithm \ref{algo:novel-representation}.

         \begin{algorithm}
            \KwIn{The waveform $x$, function $w$, and parameters $\mathbf{\Theta}=[\theta_1,\theta_2,\dots,\theta_k]$}
            \KwOut{A tensor $\bm{\mathsf{Z}}$ with $k$ channels}
            Initialize the interpolation resolutions $\omega_0$ and $n_0$ to $\infty$ \\
            \For{$i = 1$ \text{to} $k$}{
               Generate a spectrogram $\mathbf{D}_i = \text{STFT}(x; w, \theta_i)$ (Eqn \ref{eqn:stft})\\
               Maintain a running minimum of $\omega_0$ and $n_0$ \\
               \If{$\Delta\omega_i < \omega_0$}{
                  $\omega_0 = \Delta\omega_i$
               }
               \If{$\Delta n_i < n_0$}{
                  $n_0 = \Delta n_i$
               }
            }
            \For{$i = 1$ \text{to} $k$}{
               Interpolate each spectrogram $\mathbf{S}_i = \text{INTERPOLATE}(\mathbf{D}_i; \omega_0, n_0)$ (Eqn \ref{eqn:interpolation})
            }
            Stack the interpolated spectrograms $\bm{\mathsf{Z}} = [\mathbf{S}_1, \mathbf{S}_2, \dots, \mathbf{S}_k]$ \\
            Return $\bm{\mathsf{Z}}$
            \caption{Generating an instance of the novel representation}
            \label{algo:novel-representation}
         \end{algorithm}

%% Data set -------------------------
\section{Data Processing and Experiment Setup}\label{sec:dataset}
   \subsection{Recordings of Marine Mammal Vocalizations}\label{subsec:vocalizations}
\begin{figure}[H]
            \centering
               \includegraphics[width=0.7\linewidth]{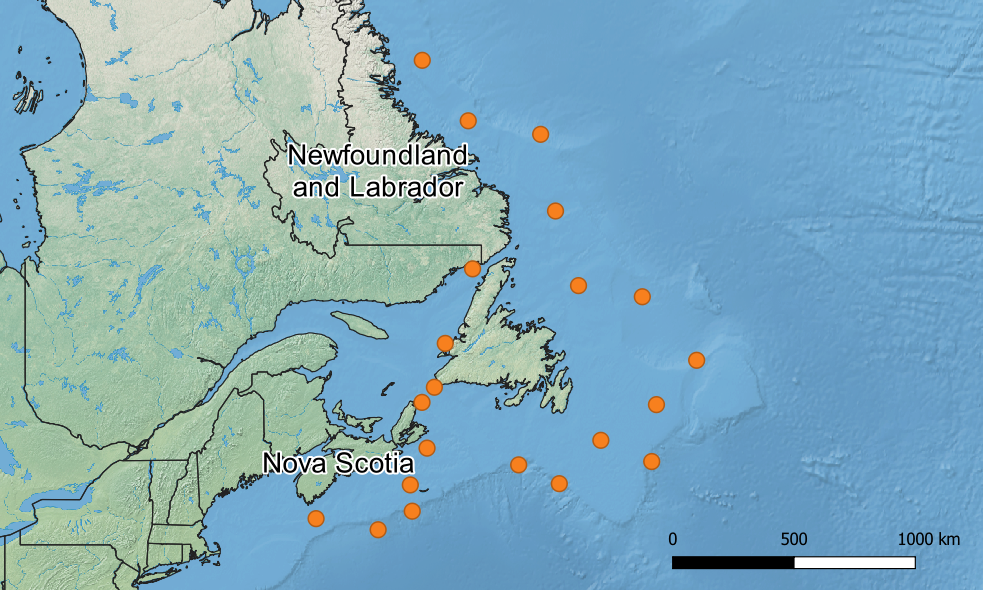}
            \caption{Map depicting the locations of the recording devices deployed by JASCO Applied Sciences along the Scotian Shelf located off the coast of Atlantic Canada.}
            \label{fig:map}
         \end{figure}

      The acoustic recordings used to train the classifier were collected by JASCO Applied Sciences using Autonomous Multichannel Acoustic Recorders (AMARs) during the summer and fall months of 2015 and 2016 in the areas surrounding the Scotian Shelf; along the coast of the Atlantic Canadian provinces (Figure \ref{fig:map}).

      The recordings were sampled at both 8kHz and 250kHz in order to capture the low frequency vocalizations of baleen whales and high frequency vocalizations of toothed whales, respectively. In this work we focus on the detection and classification of baleen whales. In particular, we are interested in three species: blue whales (\textit{Balaenoptera musculus}), fin whales (\textit{Balaenoptera physalus}), and sei whales (\textit{Balaenoptera borealis}). These species can be particularly challenging to classify as they are each capable of making a similar vocalization known as a down sweep during the summer months. A large collection of baleen whale vocalizations fall below 1000Hz, therefore, we restrict our set of acoustic recordings to those collected using the 8kHz sampling rate.

      The acoustic recordings were analyzed by marine biology experts producing over \num{30000} annotations in the form of bounding boxes around signals pertaining to the three species of whales and other acoustic sources labelled as ``non-biological''. Other species of whales present in the recording area were also annotated, however, they were not included in this paper. The distribution of annotations is heavily unbalanced in favour of the more vocal fin whales at a 6:1 ratio.

\begin{figure}[!htb]
            \centering
            \includegraphics[width=0.8\linewidth]{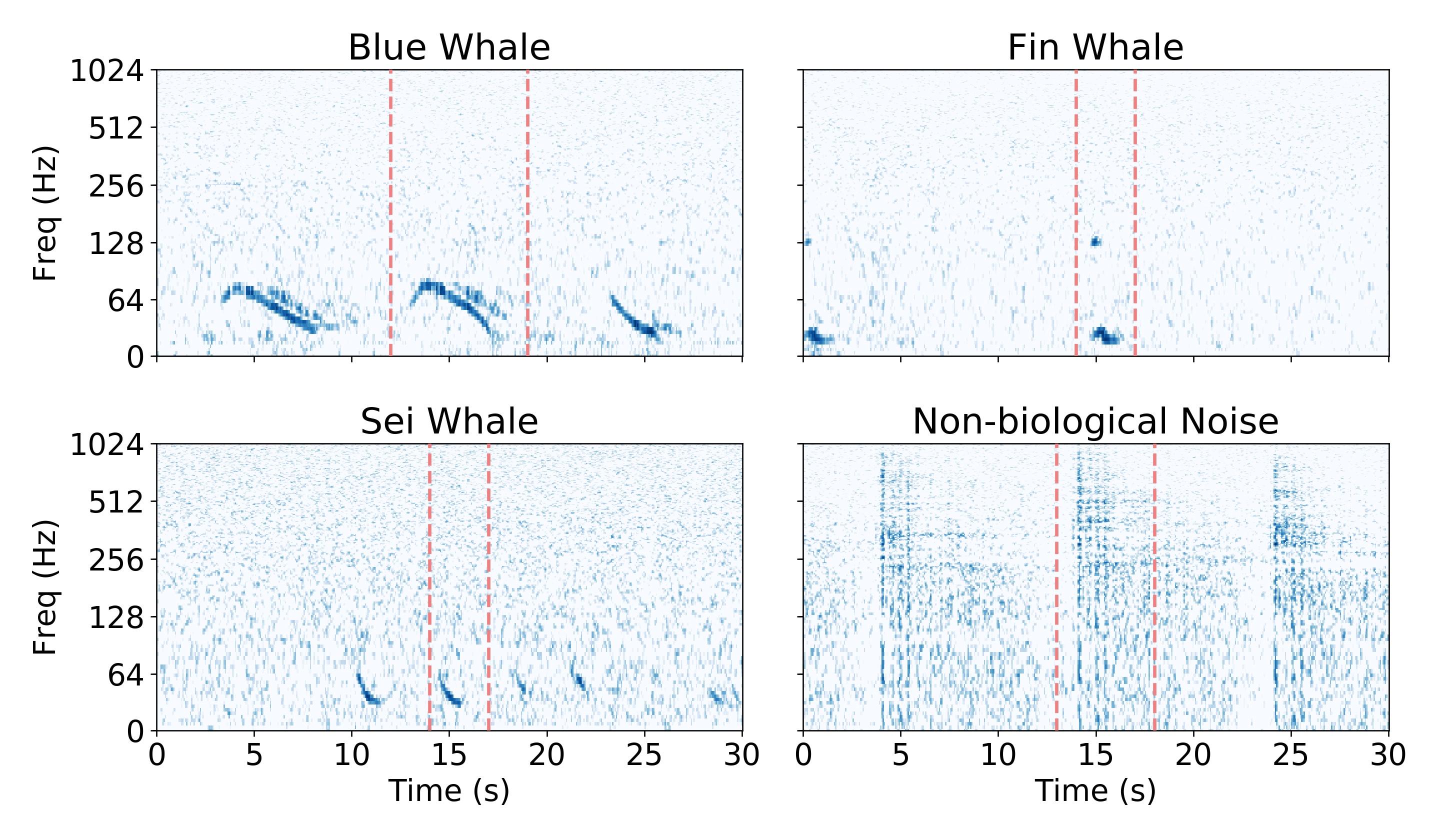}
            \caption{Example spectrograms displaying frequency in hertz on a log-scale. Examples are provided for each of the three whale species: a) blue whales, b) fin whales, c) sei whales, and d) non-biological noise. Dashed vertical lines depict the upper and lower bounds of the expert annotations.}
            \label{fig:example-spectrograms}
         \end{figure}

      The data sets used for training, validating, and testing each classifier were created in the following fashion. First, the human annotations were centred within an excerpt 30 seconds long. Four spectrograms depicting typical examples of the 30 second excerpts are provided in Figure \ref{fig:example-spectrograms}; one for each of the possible acoustic sources. Example annotations are drawn using dashed vertical lines. As we can see, not every vocalization that appeared in a spectrogram was labelled. In Figure \ref{fig:example-spectrograms}a for example, there appears to be three blue whale vocalizations occurring consecutively, however, only the second has been annotated.

      For each 30 second excerpt, a smaller ten second long sample (here on referred to as simply a ``sample'') containing the annotation was randomly selected from the larger excerpt. Due to the partial labelling of the recordings, it is possible that a sample may include more than one vocalization. For example, a sample from time 10 to 20 seconds in the file used to produce Figure \ref{fig:example-spectrograms}c, would in fact contain three sei whale vocalizations. The set of data containing only ambient noise was produced in a similar fashion, however, they were produced from a large set of files that were known to not contain baleen whale vocalizations. As such, the sampling routine simply selected a ten second sample randomly from the entire file.

      A spectrogram of each sample was produced corresponding to the CNN that was being trained and the matrices corresponding to the values of the spectrograms were used as training instances. In total there were five categories of classifiers: three trained on single-channel spectrograms using increasing FFT window lengths (i.e., \num{256}, \num{2048}, and \num{16384} samples); one trained on single-channel mel-scaled spectrograms using a window length of \num{2048} samples and 128 mels; and one trained on a three-channel version of the novel representation described in Section \ref{subsec:interpolated}. The three spectrograms used in creating the novel representation used window lengths of \num{256}, \num{2048}, and \num{16384} samples respectively and were interpolated to fit within a grid of height 256 and width 128 units. All of the above spectrograms were produced using the Hann window function and used an FFT window overlap of $1/4$ the window length. The choice of window lengths were chosen in order to capture short sweeping vocalizations such as whistles (i.e., $256\approx1/32$ the sampling rate), a more inclusive group of vocalizations (i.e., $2048\approx1/4$ the sampling rate), and long vocalizations that are fairly persistent in frequency (i.e., $16384\approx2\times$ the sampling rate). The computed spectrograms were truncated using an upper frequency bound of 1000Hz and a lower bound of 10Hz. Apart from the linear interpolant applied in the case of the novel representation, no additional filtering, smoothing, or noise removal was applied to the spectrograms.

      In practice, the ten second sampling routine and all subsequent steps including spectrogram generation were executed in parallel on the CPU while the CNN was trained on the GPU. In this way, the sampling routine acted as a quasi-data-augmentation strategy for each training batch. Further details with respect to the CPU, GPU, batch sizes, and other parameters used during training are provided in Section \ref{subsec:training-details}.

      Separate training, validation, and test data sets were produced using a random split ratio of 70/15/15, respectively. Table \ref{tbl:datasets} contains the number of files and the corresponding species distributions of each data set.

         \begin{table}[!htb]
            \caption{Number of files and the distribution of each acoustic source for the training, validation, and test sets.}
            \label{tbl:datasets}
            \centering
            \def\arraystretch{1.25}
            \begin{tabular}{lcccrlcrlcrl}
               \specialrule{.1em}{.25em}{.05em}
               \textbf{Source} & ~ & \textbf{Label} & ~~~ & \multicolumn{2}{c}{\textbf{Training}} & ~~~ & \multicolumn{2}{c}{\textbf{Validation}} & ~~~ & \multicolumn{2}{c}{\textbf{Test}} \\
               \hline
               Blue Whale      & ~ & BW             & ~~~ & \num{2692}   &  (6.23\%)              & ~~~ & \num{601}   & (6.49\%)                  & ~~~ & \num{574}   & (6.20\%)            \\
               Sei Whale       & ~ & SW             & ~~~ & \num{1701}   &  (3.94\%)              & ~~~ & \num{332}   & (3.59\%)                  & ~~~ & \num{383}   & (4.14\%)            \\
               Fin Whale       & ~ & FW             & ~~~ & \num{15118}  &  (35.01\%)             & ~~~ & \num{3244}  & (35.06\%)                 & ~~~ & \num{3272}  & (35.36\%)           \\
               Non-biological  & ~ & NN             & ~~~ & \num{2078}   &  (4.81\%)              & ~~~ & \num{449}   & (4.85\%)                  & ~~~ & \num{398}   & (4.30\%)            \\
               Ambient         & ~ & AB             & ~~~ & \num{21589}  &  (50.00\%)             & ~~~ & \num{4626}  & (50.00\%)                 & ~~~ & \num{4627}  & (50.00\%)           \\
               \specialrule{.1em}{.25em}{.05em}
            \end{tabular}
         \end{table}

   \subsection{Neural Architectures and Training Parameters}\label{subsec:training-details}
      We evaluate the performance of two commonly used CNN architectures, namely: ResNet-50 \cite{he2016} and VGG-19 with batch normalization \cite{simonyan2014}. The CNNs were implemented in Python using the PyTorch open source deep learning platform \cite{paszke2017}. Training was distributed over four NVIDIA P100 Pascal GPUs each equipped with 16GB of memory. The sampling routine and subsequent data processing was performed in parallel on two 12-core Intel E5-2650 CPUs.

      Each CNN--regardless of the FFT window length or number of channels--was trained using the same hyper-parameters apart from the initial learning rate, which was set to 0.001 for the ResNet architecture and 0.01 for the VGG architecture. In both cases, the learning rate decayed exponentially by a factor of 10 using a step schedule of 30 epochs. The batch size of each training step was set to 128 instances. Stochastic Gradient Descent (SGD) with momentum equal to 0.9 and weight decay equal to $1e^{-4}$ was used to optimize a cross-entropy loss function.

      The CNNs were each trained for a total of 100 epochs. After each epoch, the validation set was evaluated and the model with the best performance in terms of F-1 Score was saved. An early stopping criteria was not used, however, if the model began to overfit to the training data and the F-1 Score of the validation set did not improve, the best model with respect to the validation set was still maintained. Finally, the training process of each classifier was repeated ten times using different random number generator seeds.

%% Experimental Results -------------------------
\section{Experimental Results}\label{sec:results}
   Table \ref{tbl:results} contains the mean evaluation metrics and 95\% confidence intervals over ten training runs for the ResNet and VGG CNNs.

      \begin{table}[H]
         \centering
         \caption{Mean performance and 95\% confidence intervals of ten training/testing runs using random number generator seeds for each combination of CNN architecture and STFT parameter set.}
         \label{tbl:results}
         \def\arraystretch{1.5}
         \setlength{\tabcolsep}{4pt}
         \scriptsize
         \begin{tabular}{lccccc}
            \specialrule{.1em}{.25em}{.05em}
            \multicolumn{6}{c}{\textbf{ResNet-50 Performance}} \textbf{}\rule{0pt}{1ex} \\
            \textbf{}\rule{0pt}{2ex}                      & \multicolumn{1}{l}{\textbf{NFFT}} & \textbf{Accuracy}  & \textbf{Precision} & \textbf{Recall}    & \textbf{F-1 Score}  \\
            \hline
            \textbf{3-channels (Hz)}                      & -                                 & 0.953 ($\pm$0.016) & 0.887 ($\pm$0.045) & 0.871 ($\pm$0.036) & 0.878 ($\pm$0.031)  \\
            \multirow{3}{*}{\textbf{1-channel (Hz)}}      & 256                               & 0.883 ($\pm$0.022) & 0.714 ($\pm$0.060) & 0.641 ($\pm$0.037) & 0.675 ($\pm$0.046)  \\
                                                          & 2048                              & 0.944 ($\pm$0.009) & 0.863 ($\pm$0.036) & 0.838 ($\pm$0.039) & 0.850 ($\pm$0.023)  \\
                                                          & 16384                             & 0.943 ($\pm$0.013) & 0.860 ($\pm$0.032) & 0.847 ($\pm$0.058) & 0.853 ($\pm$0.031)  \\
            \multicolumn{1}{r}{\textbf{1-channel (mels)}} & 2048                              & 0.895 ($\pm$0.031) & 0.762 ($\pm$0.067) & 0.723 ($\pm$0.048) & 0.742 ($\pm$0.044)  \\
            \multicolumn{6}{c}{\textbf{VGG-19 Performance}} \rule{0pt}{4ex} \\
            \textbf{}\rule{0pt}{2ex}                      & \multicolumn{1}{l}{\textbf{NFFT}} & \textbf{Accuracy}  & \textbf{Precision} & \textbf{Recall}    & \textbf{F-1 Score}  \\
            \hline
            \textbf{3-channels (Hz)}                      & -                                 & 0.961 ($\pm$0.017) & 0.906 ($\pm$0.044) & 0.892 ($\pm$0.049) & 0.899 ($\pm$0.041)  \\
            \multirow{3}{*}{\textbf{1-channel (Hz)}}      & 256                               & 0.914 ($\pm$0.024) & 0.790 ($\pm$0.048) & 0.771 ($\pm$0.070) & 0.780 ($\pm$0.053)  \\
                                                          & 2048                              & 0.959 ($\pm$0.019) & 0.899 ($\pm$0.041) & 0.889 ($\pm$0.048) & 0.894 ($\pm$0.039)  \\
                                                          & 16384                             & 0.951 ($\pm$0.017) & 0.871 ($\pm$0.037) & 0.878 ($\pm$0.038) & 0.875 ($\pm$0.028)  \\
            \multicolumn{1}{r}{\textbf{1-channel (mels)}} & 2048                              & 0.918 ($\pm$0.022) & 0.818 ($\pm$0.043) & 0.784 ($\pm$0.036) & 0.801 ($\pm$0.034)  \\
            \specialrule{.1em}{.25em}{.05em}
         \end{tabular}
      \end{table}

   The classifier trained on the novel representation outperforms the remaining classifiers trained on single-channel inputs. Paired two-sample $t$          -tests indicate that the improvement in performance between the classifier trained on the novel representation is statistically significant in all cases with one exception: the VGG-19 CNN trained on single-channel inputs using a window length of 2048 samples.

\begin{figure}[H]
         \centering
         \includegraphics[width=0.675\linewidth]{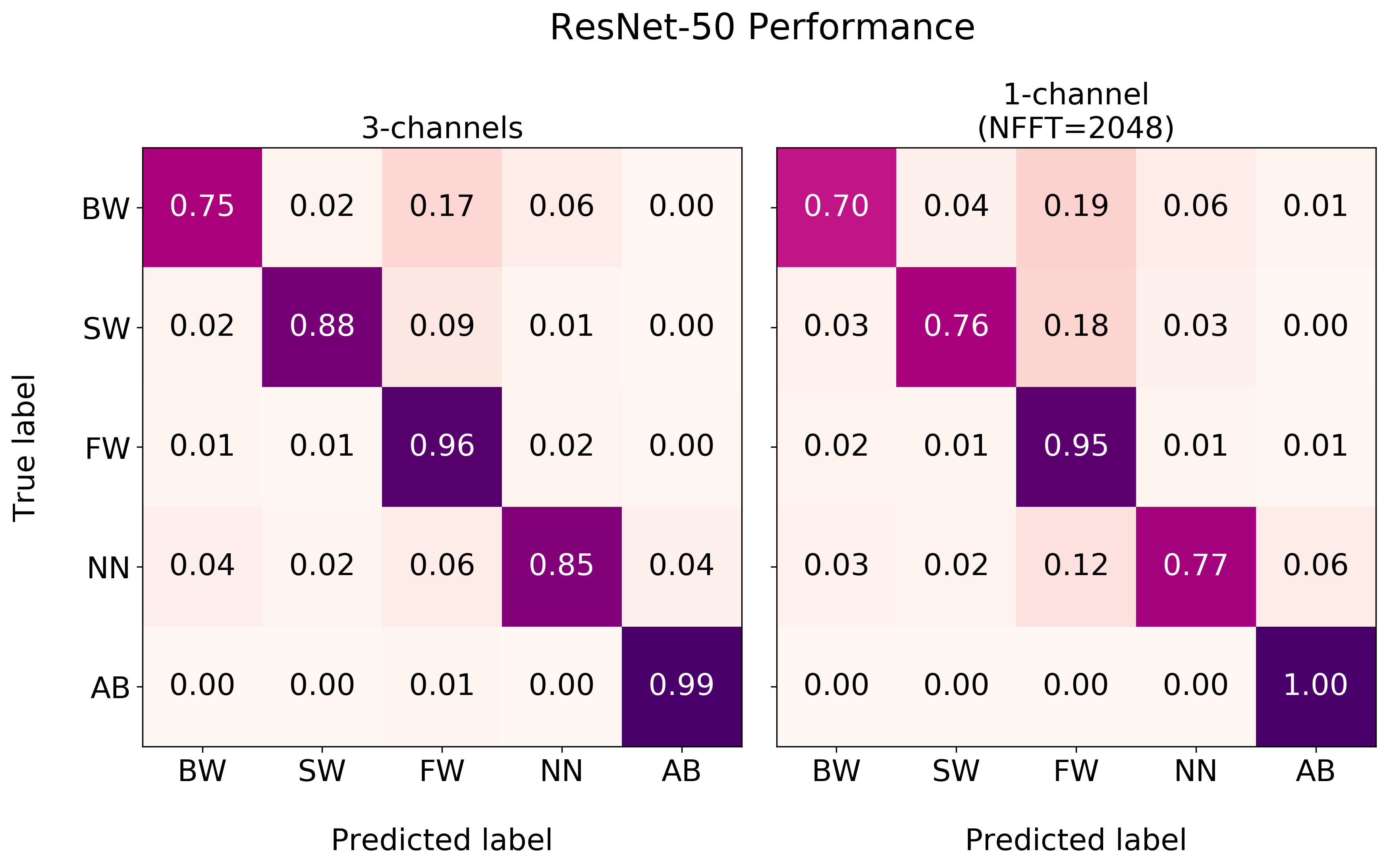}
         ~\vskip .5em
         \includegraphics[width=0.675\linewidth]{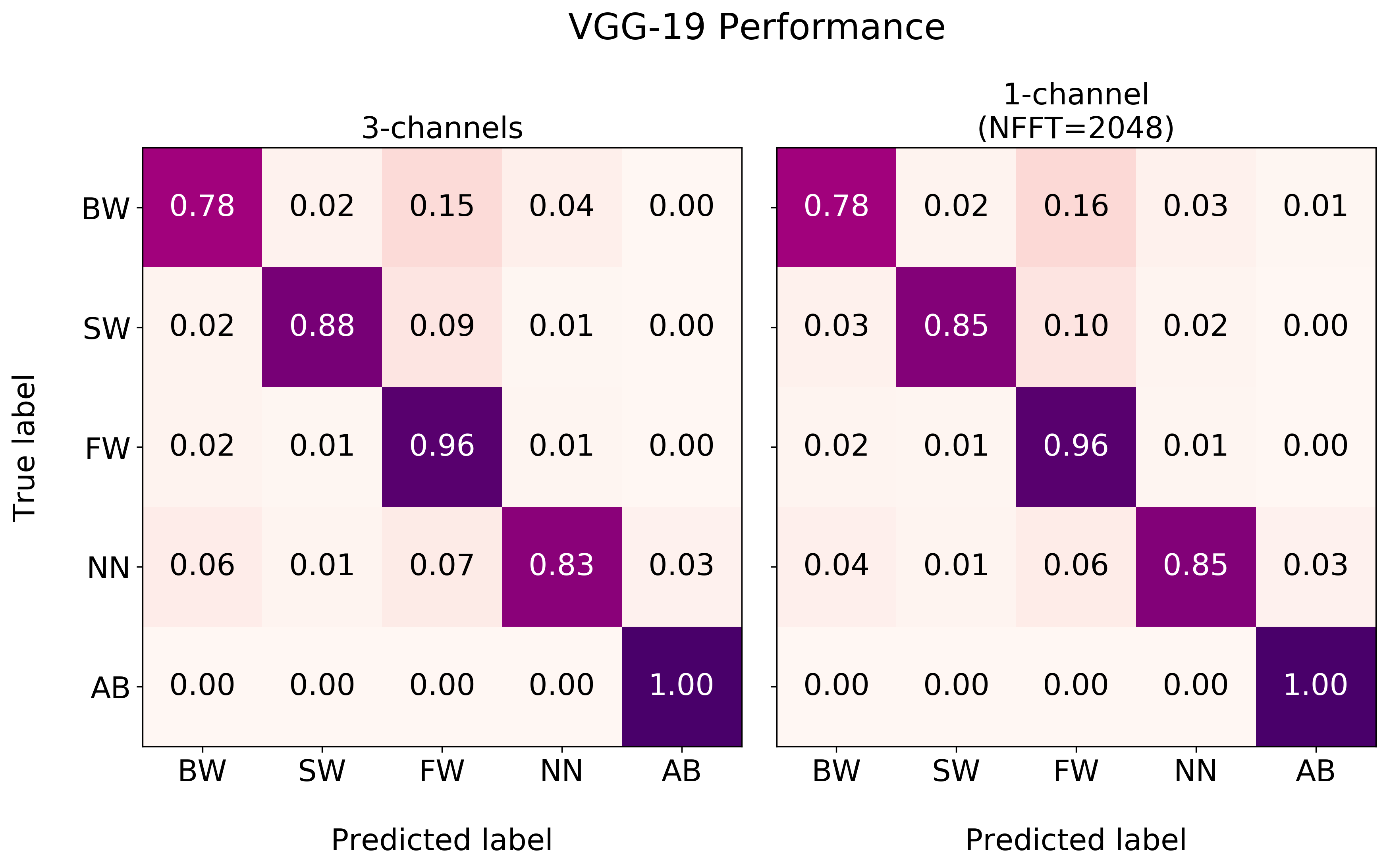}
         \caption{Normalized confusion matrices of the two best performing classifiers in terms of F-$1$ Score for the ResNet-50 and VGG-19 CNNs.}
         \label{fig:confusion-matrices}
      \end{figure}

   Figure \ref{fig:confusion-matrices} contains four confusion matrices: two corresponding to the VGG-19 architecture and two corresponding to the ResNet-50 architecture. In both cases, the best two performing classifiers were those trained on the novel representation and the single-channel linearly scaled spectrogram produced using a window length of 2048 samples.

   \subsection{Generalization to Other Acoustic Sources}\label{subsec:generalizable}
      In order to the demonstrate the ability of the DCS that we have developed to generalize to other acoustic sources below 1000Hz, we train a new classifier using a transfer learning approach to include humpback whale (\textit{Megaptera novaeangliae}) vocalizations. Specifically, all sixteen convolutional layers in the VGG-19 network trained on the novel representation are frozen. The last three layers of the network are then re-learned on the data set described in Table \ref{tbl:datasets} with an additional \num{2100} humpback vocalizations. The hyper-parameters and optimization routine used for training the last layers of the network are equivalent to those detailed in Section \ref{subsec:training-details}.

      The trained classifier achieves performance levels in terms of accuracy, precision, and recall of \num{0.948}, \num{0.884}, and \num{0.871}, respectively, without the need of re-training the convolutional feature extraction layers of the CNN.

         \begin{figure}[!htb]
            \centering
            \includegraphics[width=0.4\linewidth]{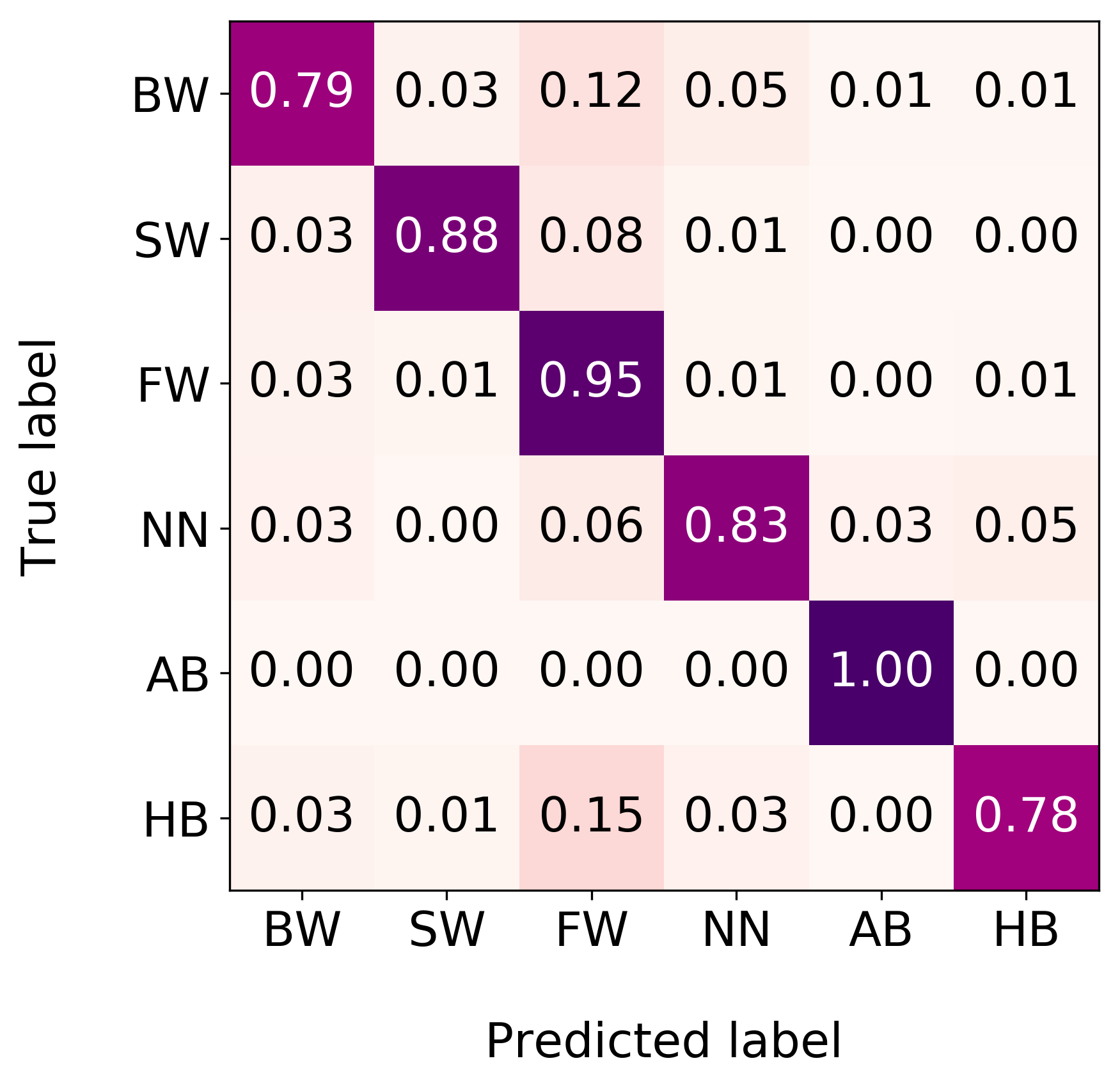}
            \caption{Normalized confusion matrix of the transfer learning experiment evaluated on the test set described in Table \ref{tbl:datasets} with an additional 450 humpback annotations identified using the label ``HB''.}
            \label{fig:humpback-confusion-matrix}
         \end{figure}

      \subsubsection{t-SNE Embeddings}\label{subsec:t-sne}
\begin{figure}[!htb]
               \centering
               \includegraphics[width=0.65\linewidth]{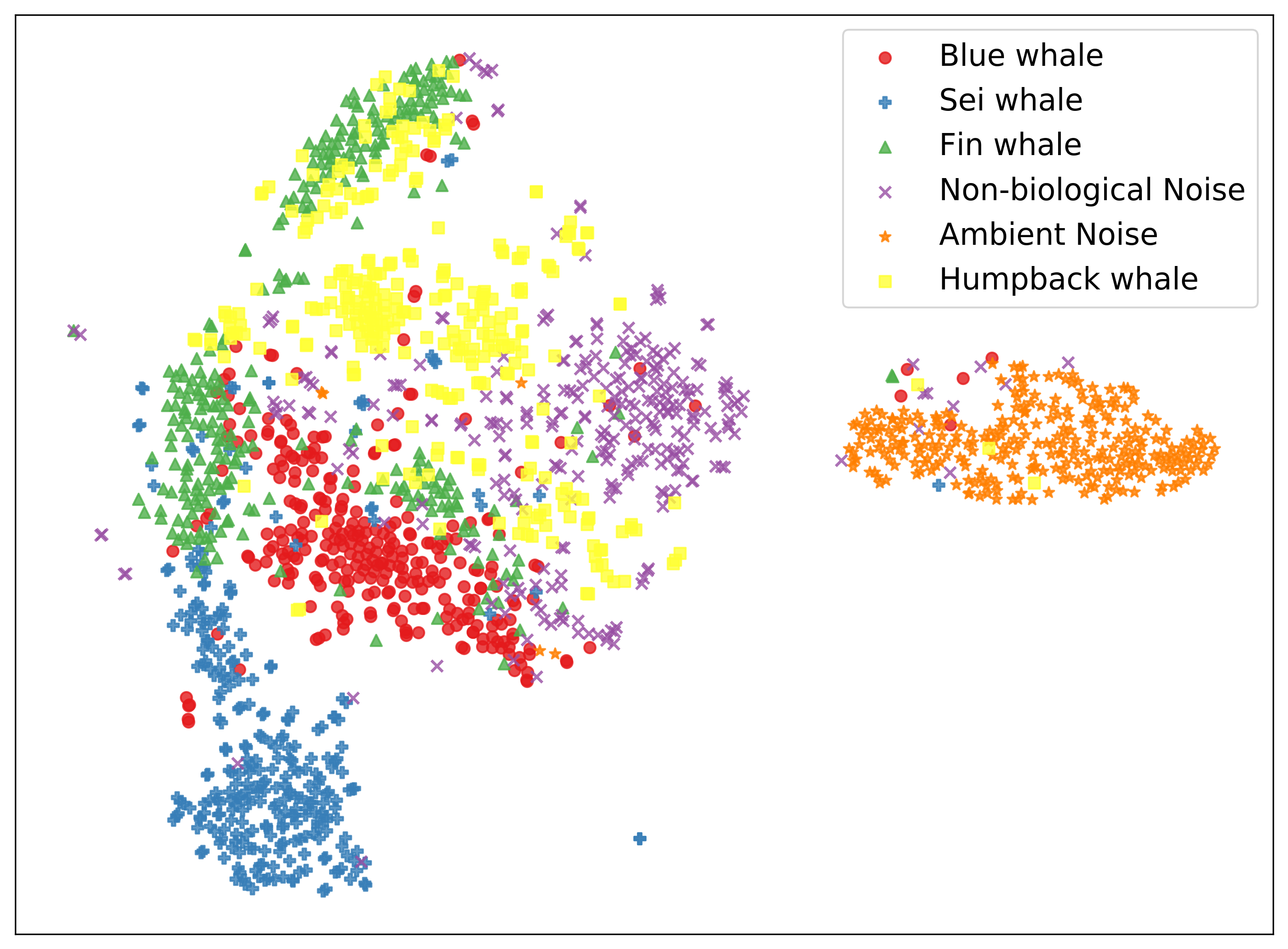}
               \caption{t-SNE embeddings computed from the output of the last frozen layer of the VGG-19 CNN architecture.}
               \label{fig:t-sne}
            \end{figure}

         The transfer learning results exhibit that the CNN is capable of learning complex features contained within a spectrogram. Further proof of this statement can be found in Figure \ref{fig:t-sne}, which contains two-dimensional t-SNE embeddings \cite{maaten2008} generated using the output of the last frozen layer of the VGG-19 CNN trained on the novel representation.

         There is a distinct separation between the original five classes of acoustic sources. More importantly, even before learning the last three classifying layers of the VGG-19 CNN, a relatively distinct representation has already been learned for the humpback whale class. This result is significant as it implies additional species with less annotated data can be included in our implementation of a DCS through transfer learning.

%% Conclusion -------------------------
\section{Conclusion}\label{sec:conclusion}
   This paper presents a scientific application focused on detecting and classifying marine mammal vocalizations in acoustic recordings. In particular, we have developed a DCS based on Convolutional Neural Networks that is capable of classifying three species of baleen whales as well as non-biological sources against a background of ambient noise. A novel representation of acoustic signals was introduced and this representation increased the performance of the aforementioned classifier. The DCS was shown to be capable of learning generalizable representations for the purpose of including additional species. The latter note is substantial as it implies that species with very little annotated data--especially those species that are endangered--can be included in the training process of future classifiers through transfer learning.

   A well performing and generalizable DCS such as the one that we have developed is of great interest to researchers in the fields of marine biology, bioacoustics, and oceanography as it allows for fast analysis of large acoustic data sets. Such analysis may be used to inform governmental and environmental policy pertaining to conservation efforts of marine mammal life.

   \subsection{Future Work}\label{subsec:future-work}
      The work presented above is part of an ongoing research project focused on developing a DCS to be used in real time on specially developed autonomous hardware (e.g., moored recording devices and/or ocean gliders). With this goal in mind, we must consider time/space complexities and additional research into model compression is necessary. Further research and development is ongoing using data collected from recording devices deployed in various locations around the world. The supplementary data allows for the ability to include a variety of additional species of baleen whales as well as other marine mammals (e.g., pinnipeds). The additional data will also allow for the interpretation of different sources of ambient noise (i.e., soundscapes). Collectively, including additional data from various locations around the world will lead to a more robust DCS of marine mammal vocalizations.

      Another option for including additional species of marine mammals for which we have little available data is through data augmentation strategies. In particular, research into using unsupervised or semi-supervised approaches (e.g., Variational Auto-encoders, Generative Adversarial Networks) to increase the size of the training data could be highly beneficial.

      Recent work into neural network architectures that operate directly on the waveform of an acoustic signal have shown great promise \cite{van2016}. While the majority of these results are specific to generative tasks, these architectures--or a suitable alternative--may be used in training a classifier for acoustic recordings such as those described in this paper. In particular, through learning from the waveform directly we avoid any information loss that takes place during a Fourier transform.

      Finally, given the promising results of our early experiments reported in Section \ref{subsec:generalizable}, we also plan on investigating the use of various transfer learning and meta-learning techniques for the task at hand.

%% Acknowledgements -------------------------
\section*{Acknowledgements}
   Collaboration between researchers at JASCO Applied Sciences and Dalhousie University was made possible through a Natural Sciences and Engineering Research Council Engage Grant. The acoustic recordings described in this paper were collected by JASCO Applied Sciences under a contribution agreement with the Environmental Studies Research Fund.

%% References -------------------------
\bibliographystyle{splncs04}

\end{document}